\documentstyle[11pt,paspconf]{article}

\def\apjs#1{{\em Astrophys. J. Suppl.} {\bf #1}}
\def\be{\begin{equation}}
\def\ee{\end{equation}}
\def\bea{\begin{eqnarray}}
\def\eea{\end{eqnarray}}
\def\etal{{\it et al.}\ }

\begin{document}

\title{Structure and Evolution of Stellar Populations in Local Group
Galaxies}

\author{J. Einasto, P. Tenjes}

\affil{Tartu Observatory, EE-2444 T\~oravere, Estonia}

%\maketitle

\begin{abstract}
We use detailed modeling of stellar populations of nearby galaxies to
calculate population parameters. Our sample of galaxies includes most
galaxies of the Local Group and several more distant giant galaxies.
Available data are sufficient to decompose galaxies into the nucleus,
the metal rich core, the bulge, the halo, the young and old disks, and
the dark halo. We compare mass-to-luminosity ratios and colors of
bulges, disks and halos with results of models of galactic chemical
evolution. The luminosity weighted mean of the mass-to-luminosity
ratio of visible stellar population is $M/L_B = 4 \pm 1.4$ in solar
units. This ratio is surprisingly constant for galaxies of very
different absolute magnitude.
\end{abstract}

\keywords{galaxies, observation, models, chemical evolution}

\section{Introduction}

Direct photometric, spectroscopic and dynamical observations yield
integrated information on galaxies. To find physical parameters of
individual galactic populations detailed modeling of galaxies and
their populations is needed. In most models only 3 luminous
populations are considered, the bulge, the halo and the disk, and
additionally the dark halo.  A method to determine population
parameters from observational data has been elaborated by Einasto
(1974) and Einasto \& Haud (1989). This method permits to decompose
galaxies into more stellar populations. First series of models was
reported on the First European Astronomical Meeting (Einasto 1974). A
new series of models with better data and improved modeling procedure
is now in progress, so far models are published for our Galaxy (Haud \&
Einasto 1989), M87 (Tenjes, Einasto \& Haud 1991), M31 (Tenjes, Haud
\& Einasto 1994), M81 (Tenjes, Haud \& Einasto 1998). Modeling
information is presently available for most members of the Local Group
and several other galaxies. Data for more distant bright galaxies is
useful to compare population parameters for galaxies of different
morphological types.

Here we shall review principal modeling data for all galaxies for
which detailed models are available. We consider only three principal
populations of galaxies, the bulge, halo and disk, data on other
populations can be found in original publications. We compare
population parameters with results of calculation of chemical
evolution of galaxies.

\section{Population models}

In our models we assume that galaxies are in hydrostatic equilibrium,
i.e.  that rotational velocities of young disk objects can be
identified with circular velocities. We describe the spatial density
distribution of populations by the modified exponential law:
$$
\rho(a) = \rho(0) \exp(-[a/(k a_0)]^{1/N}).
$$
Here $\rho(0)$ is the central density of the population, $a$ is the
major semi-axis of the equidensity ellipsoid, $a_0$ is the harmonic
mean radius of the population, and $k$ and $N$ are dimensionless
structural parameters. The parameter $N$ determines the density law of
the population; $N=0.5$, $N=1$ and $N=4$ correspond to the
conventional Gaussian, exponential, and de Vaucouleurs models (de
Vaucouleurs applied this law for projected density, here we use it for
the spatial density).

\begin{table*}
\caption{Galaxy parameters}
\label{tab:param}
\begin{center}\tiny
\begin{tabular}{lccccccccccc}
\hline\hline\\
Name & & Type & \multicolumn{2}{c|}{Disk} & \multicolumn{2}{c|}{Bulge}
& \multicolumn{2}{c|}{Halo} & \multicolumn{2}{c|}{Visible} & Total\\
 & $M_B$ &  & \multicolumn{2}{c|}{ } & \multicolumn{2}{c|}{ } &
\multicolumn{2}{c|}{ } & \multicolumn{2}{c|}{matter} &  \\
\cline{4-12}\\
 & & & $M/L_B$ & $B-V$ & $M/L_B$ & $B-V$ & $M/L_B$ & $B-V$ & $M/L_B$ &
$B-V$ & $M/L_B$ \\
\\
\hline
\\
M31     &--20.8&Sb  &7.1&0.71&3.0&0.98&2.0&0.79&4.8&0.80&200   \\
M32     &--15.7&cE2 &   &    &2.4&0.90&   &    &2.4&0.90&      \\
M81     &--20.2&Sab &9.4&0.76&5.8&1.02&2.4&0.70&5.4&0.83&230   \\
M87     &--22.2&E0-1&   &    &8.0&1.01&2.5&0.74&4.9&0.87&1200:\\
M104    &--22.3&Sa  &5.0&1.00&2.9&1.01&2.7&0.77&3.2&0.87&47:  \\
        &      &    &   &    &   &    &   &    &   &    &      \\
NGC 3115&--20.0&S0  &6.7&0.93&4.9&1.00&2.3&0.75&4.6&0.87&180   \\
NGC 2841&--20.4&Sb  &9.3&0.95&7.7&0.95&   &    &8.9&0.95&220   \\
NGC 4321&--20.7&Sbc &3.3&0.74&1.7&0.60&   &    &3.1&0.71&?     \\
NGC 5457&--21.3&Scd &3.2&0.62&3.4&0.78&   &    &3.3&0.62&150   \\
NGC 6503&--18.0&Scd &2.0&0.71&1.8&0.57&   &    &2.0&0.71&130: \\
        &      &    &   &    &   &    &   &    &   &    &      \\
NGC 7793&--18.7&Sdm &2.5&0.55&2.0&0.62&   &    &2.5&0.55&300: \\
Fornax  &--12.1&dSph&   &    &   &    &1.2&0.63&1.2&0.63&$>12$\\
Sculptor&--10.1&dSph&   &    &   &    &2.0&0.67&2.0&0.67&$>8$ \\
Carina  & --8.0&dSph&   &    &   &    &1.3&0.70&1.3&0.70&$>40$\\
Draco   & --8.1&dSph&   &    &   &    &1.9&0.62&1.9&0.62&$>80$\\
        &      &    &   &    &   &    &   &    &   &    &      \\
U.Min   & --8.3&dSph&   &    &   &    &1.5&0.63&1.5&0.63&$>60$\\
Leo I   &--10.3&dSph&   &    &   &    &1.4&0.80&1.4&0.80&$>7$ \\
Sextans & --8.4&dSph&   &    &   &    &1.6&0.66&1.6&0.66&$>120$\\
Leo II  & --8.6&dSph&   &    &   &    &1.7&0.65&1.7&0.65&$>10$\\
\\
\hline\hline
\end{tabular}
\end{center}
\end{table*}
\normalsize

\begin{figure}
\vspace*{9cm}
\caption{Population parameters for galactic disks, bulges and halos, and
for all visible matter.
}
\includegraphics{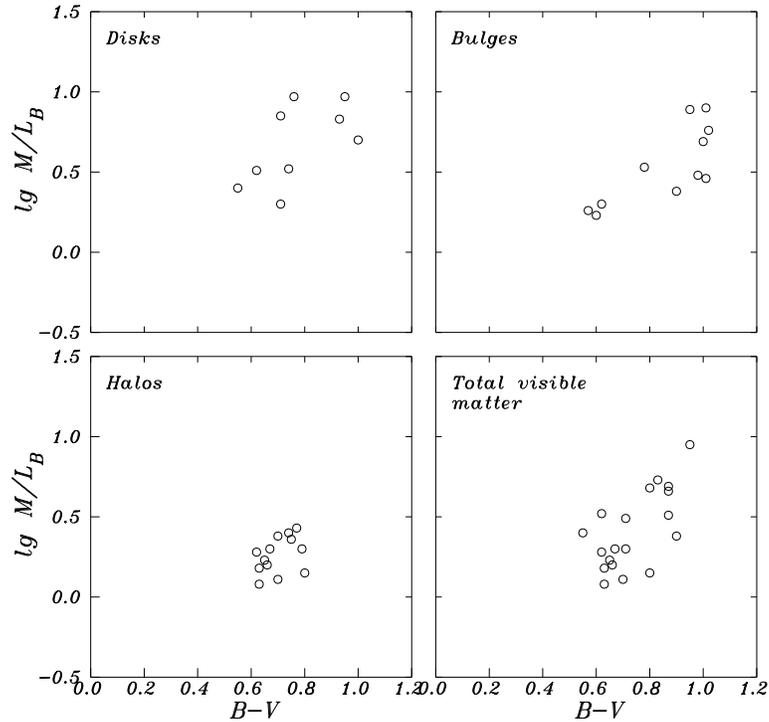}
\label{figure1}
\end{figure}

\begin{figure}[h]
\vspace*{6cm}
\caption{Comparison of parameters of galactic populations with models
of chemical evolution. Left panel shows integrated data on galaxies,
right panel data on galactic populations.  Galaxies are denoted as
follows: open circles -- E/S0, Sa, Sb galaxies; triangles -- Sc and Sd
galaxies.  Chemical evolution models are given for galaxies of
different morphological type (solid lines); dotted lines show
isochrones.  Evolution models are taken from Tantalo \etal (1998). In
right panel open circles are for bulges, stars for globular clusters,
squares for halos.  Solid lines show evolution models for single
populations of metallicities [Fe/H]~$= -2, ~-1, ~-0.3, ~0, ~0.4$ (from
left to right) and ages from 1.5 to 17 Gyrs (from bottom up) (Worthey
1994, 1996). For metallicities [Fe/H]~$/leq -0.3$ only ages greater
than 8 Gyrs are available.  } 
\includegraphics{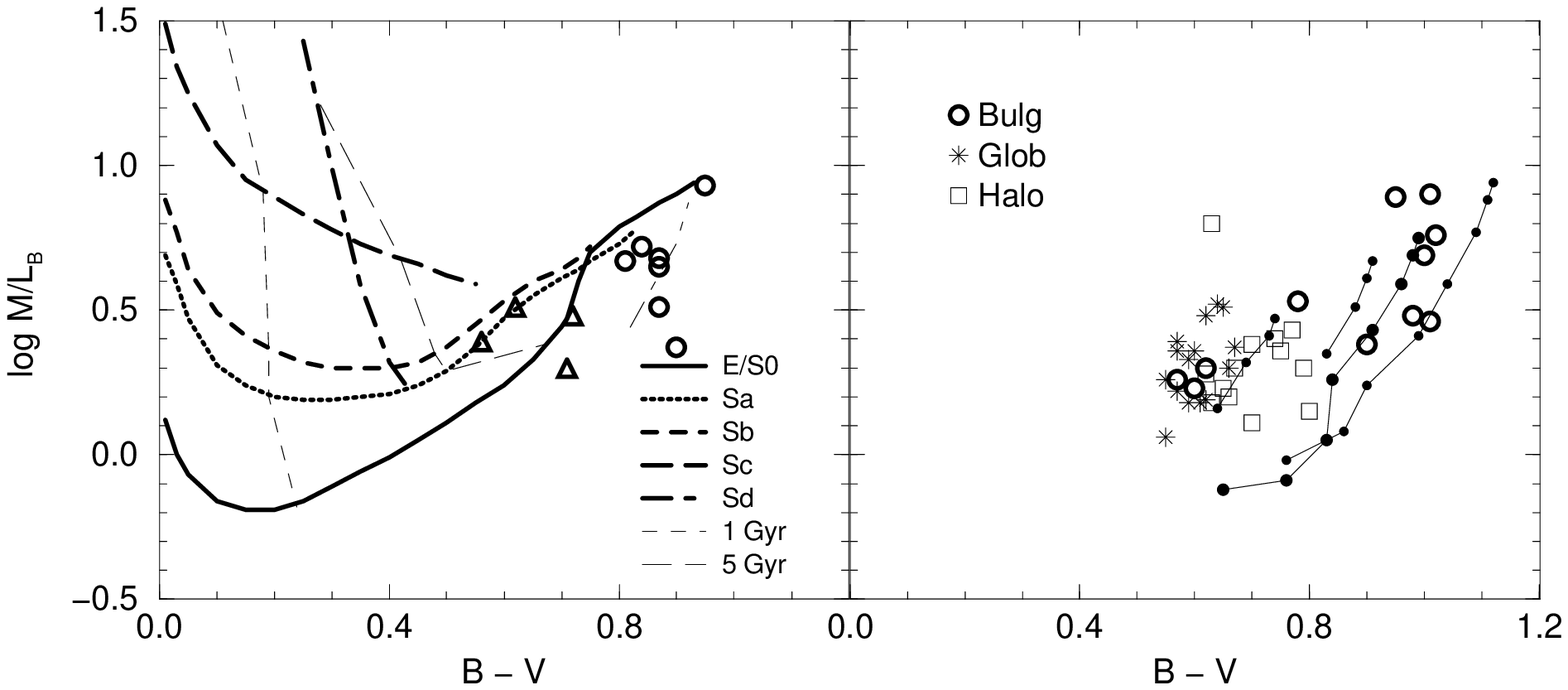}
\label{figure2}
\end{figure}

Population models are found by successive approximations. First, a
zero approximation model is determined from fits of observed
descriptive functions with standard functions (density law, rotational
velocity law etc.). Using approximate values for population parameters
model descriptive functions are calculated and compared with
observations. The difference between calculated and observed functions
gives information to correct population parameters; and the next
approximation is found.  The program looks, which population
parameters influence descriptive functions less, these parameters are
temporarily excluded from the least squares fit. Our experience shows
that the process converges rapidly and population parameters found are
rather stable, i.e. small deviations in observed functions do not lead
to large deviations of population parameters.

\section{Parameters of stellar populations}

Table 1 gives the summary of main population parameters for our
models. We give here only two parameters for each population, the
mass-to-luminosity ratio, $M/L_B$, and the color index, $B -
V$. Additionally we give the mass-to-luminosity ratio and the color
for the whole galaxy (for visible populations), $M/L_B$ is given also
for the galaxy as a whole, including its dark halo.  These parameters
characterize the mean stellar content and the evolutionary status of
the galaxy.

Population parameters are plotted in Figure~1, separately for disks,
bulges, and halos, and for the total visible matter. We see that
parameters of disks and bulges have a fairly large spread, whereas the
spread of parameters for halos is much smaller. Evidently, halos are
more homogeneous in their stellar content. There exists a weak
relation between the mean color and the mass-to-luminosity ratio for
disks, bulges and total visible matter: redder populations have higher
mass-to-luminosity ratios.

The relation between color and $M/L_B$ is easily explained by theoretical
models, see Figure~2. In young populations luminous blue main sequence
stars dominate, in old populations dominating stars are red giants which
are less luminous and red.  The agreement of evolution models with
population models is very good quantitatively. This agreement shows that
current models of chemical evolution are accurate enough, and that 
our population models describe actual properties of populations
satisfactorily.

\begin{figure}[h]
\vspace*{6cm}
\caption{Mass-to-luminosity ratio of galactic populations in galaxies of
different total absolute magnitude, $M_B$. Open circles denote bulges,
asterisks -- halos, dots -- disks, open squares denote total visible
matter in galaxies. 
} 
\includegraphics{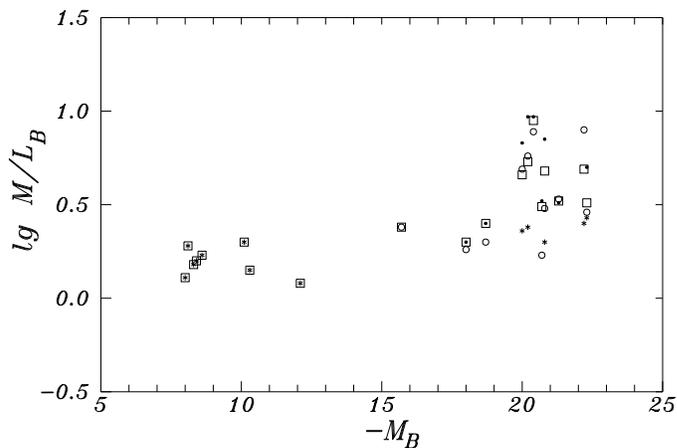}
\label{figure3}
\end{figure}

It is interesting to compare mean $M/L_B$ of galaxies with their absolute
magnitudes. This is done in Figure~3. Our models span a very large range
in absolute magnitude -- dwarf spheroidal galaxies are more than million
times less luminous than giant spiral and elliptical galaxies. The figure
shows, in a different perspective, that stellar halos have rather small
spread of their properties. For disks and bulges the spread is larger, but
also rather limited: all populations have $M/L_B$ between 1 and 10. 

The mean mass-to-luminosity ratio for all visible matter, weighed with the
luminosities of galaxies, is $M/L_B= 4.1 \pm 1.4$ for the luminosity
function of Sandage \etal (1985).

\section{Discussion and summary}

The first set of population models was reported on the First European
Astronomical Meeting in Athens 1972. At this time models for our
Galaxy, M31, M32, M87, Fornax and Sculptor galaxy were available.
This first application of the modeling method had several important
consequences. It was clear that galaxies are surrounded by dark halos
or coronas (Einasto 1974). Further it was found that all stellar
populations have a mass-to-luminosity ration between 1 and $\approx
10$ (except very metal-rich nuclei and cores which have slightly
higher $M/L_B$ values).  Remember that in early 70's most astronomers
believed that $M/L_B$ of bulges of elliptical galaxies are very high
(to explain large relative velocities of their companion galaxies).
Oldest metal poor population, the stellar halo, has the lowest $M/L_B$
of all populations.  On the other hand, the extended dark population
has extremely high $M/L_B$, and thus differs very much from all
conventional stellar populations.  A few years later a direct
confirmation of a low $M/L_B$ for bulges of elliptical galaxies
followed when Faber \etal (1977) measured the rotational velocity of
the thin gaseous disk of the Sombrero galaxy and derived for the bulge
of this galaxy $M/L_B \approx 3$, in very good agreement with our
population models for bulges of other galaxies.

Modern modeling data have confirmed these early results. The basic
difference between old and new models lies in the fact, that,
according to new data, velocity dispersions near the centers of
galaxies are lower than believed earlier, which leads to smaller
values of $M/L_B$. Thus the spread of $M/L_B$ of stellar populations
is, according to new data, smaller than assumed in 70's.  The contrast
between ordinary populations and the dark population is even
greater. We repeat here main differences of parameters of stellar
populations and the dark matter halo.

1. All stellar populations have $1 \leq M/L_B \leq  10$, the dark
matter population has $M/L_B >> 1000$.

2. There exist a continuous transition of stellar populations from
stellar halo to bulge, from bulge to old disk, from old to young disk;
these intermediate populations are observed in our Galaxy. All stellar
populations contain a continuous sequence of stars of different mass,
some of these stars have masses corresponding to the mass of red
giants for the age of the population.  Red giants have high
luminosity, thus all stellar populations are visible, in contrast to
the dark population.

3. The density of stellar populations rapidly increases towards the plane
or the center of the galaxy, the dark matter population has very low
central concentration of mass.

These arguments show qualitatively that dark matter must have been
originated much earlier than ordinary stellar populations, and that
there must be a large gap between the formation time of the dark halo
and oldest visible stellar populations, since there exist no
intermediate population between the dark and ordinary populations.
These arguments were known already in early 70's and played
crucial role in the development of the concept of dark matter in the
Universe. 

In the present set of models parameters of stellar populations are
fixed rather well (for the discussion of parameter errors see original
papers).  The central density of the dark halo is also fixed with good
accuracy. The least accurate parameter is the radius of the dark halo
and its total mass, since there are only few mass indicators on large
distances from the main galaxy.  Local Group offers a unique
possibility to determine the mass and the radius of dark halos of the
Galaxy and M31 with high accuracy.  The level of the rotational
velocity of both galaxies is known well, this determines the density
in inner regions of dark halos.  The total mass of galaxies is given
by the relative velocity of centers of galaxies, following arguments
by Kahn and Woltjer (1959), Einasto and Lynden-Bell (1982) and
Valtonen \etal (1993). This calculations shows that masses determined
from internal dynamics of galaxies and their relative velocity
coincide if we adopt outer radii of dark halos about 200 -- 300 kpc.

\end{document}